\documentclass[article,table,nojss]{jss}

\usepackage{orcidlink,thumbpdf,lmodern}

\usepackage{amsmath, amssymb}
\usepackage[ruled, linesnumbered]{algorithm2e}
\usepackage{tikz}
\usepackage{float}
\newcommand{\onto}{\twoheadrightarrow}
\newcommand\xonto[2][]{%
  \mathrel{\ooalign{$\xrightarrow[#1\mkern4mu]{#2\mkern4mu}$\cr%
  \hidewidth$\rightarrow\mkern4mu$}}
}

\newcommand{\class}[1]{`\code{#1}'}

\newcommand{\xfnm}[1][]{\ifx!#1!\else\unskip,\space#1\fi}

\author{Mark P.J. van der Loo~\orcidlink{0000-0002-9807-4686}\\
        Statistics Netherlands and Leiden University}

\Plainauthor{Mark P.J. van der Loo}

\title{Split-Apply-Combine with Dynamic Grouping}
\Plaintitle{Split-Apply-Combine with Dynamic Grouping}
\Shorttitle{Split-Apply-Combine with Dynamic Grouping}

\Abstract{
Partitioning a data set by one or more of its attributes and computing an
aggregate for each part is one of the most common operations in data analyses.
There are use cases where the partitioning is determined dynamically by
collapsing smaller subsets into larger ones, to ensure sufficient support for
the computed aggregate. These use cases are not supported by software
implementing split-apply-combine types of operations. This paper presents the
\proglang{R} package \code{accumulate} that offers convenient interfaces for
defining grouped aggregation where the grouping itself is dynamically
determined, based on user-defined conditions on subsets, and a user-defined
subset collapsing scheme. The formal underlaying algorithm is described and
analyzed as well.
}

\Plainkeywords{data analysis, R}

\Address{
  Mark P.J. van der Loo~\orcidlink{0000-0002-9807-4686}\\
  Research and Development\\
  Statistics Netherlands\\
  Henri Faasdreef 312\\
  2492JP Den Haag, the Netherlands\\
  E-mail: \email{mpj.vanderloo@cbs.nl}\\
  URL: \url{https://www.markvanderloo.eu}\\
  \emph{and}\\
  Leiden Institute of Advanced Computer Science (LIACS)\\
  Leiden University\\
  P.O. Box 9512\\
  2300 RA Leiden, The Netherlands\\
}

\begin{document}



\section{Introduction}
The operation of splitting a data set into non-overlapping groups, computing an
aggregate for each group, and combining the results into a new dataset is one
of the most common operations in data analyses. Indeed, any software for data
analyses includes some functionality for this. For example, the combination of
\code{split}/\code{lapply}/\code{unsplit} as well as \code{aggregate} have been
a part of the \proglang{S} \citep{becker1988new} and \proglang{R} \citep{rcore}
languages for a long time. For \proglang{R} there are several packages that
implement functionality for this, including \pkg{plyr}
\citep{wickham2011split}, it's successor \pkg{dplyr} \citep{wickham2022dplyr},
it's drop-in replacement \pkg{poorman} \citep{eastwood2022poorman}, and
performance-focused \proglang{R} packages \pkg{collapse}
\citep{krantz2022collapse} and \pkg{data.table} \citep{dowle2022datatable}. In
\proglang{Python} the \code{pandas} package implements several methods for
grouping records and aggregating over one or more columns in data frame
objects. The more recent \citet{polars2023} library for \proglang{Python} and
\proglang{Rust} also implement such features. Similarly, the \pkg{DataFrames}
package for \proglang{julia} implements split-apply-combine functionality
\citep{kaminski2022dataframes}.

In all packages mentioned, the calculation for each group uses data available
within the group. However, there are valid use cases where a group aggregate is
determined using attributes from out-of-group entities. One example where this
occurs is in the area of Small Area Estimation (SAE, see \emph{e.g.},
\citet{rao2015small, molina2015sae}). Here, one wishes to estimate an aggregate
for a group, for example a geographical region, or a detailed population
subset, where the number of (sampled) observations is so small that the variance
of the estimate would be unacceptably large. In small area estimation (SAE) one
trades bias for variance by `borrowing statistical strength' from out-of-group
records. The out-of-group records can be obtained, for example by combining the
original small group with a group of records that are deemed similar in certain
respects. A second area where out-of-group records play a role is in certain
hot-deck imputation methods \citep{andridge2010review}.  In
$k$-nearest-neighbours imputation for example, one finds a set of $k$ donor
records that are preferably in the same group, but this condition may be
relaxed if there are not enough records in the group.  In the \pkg{VIM} package
for \proglang{R} \citep{kowarik2016imputation}, this is controlled by a
combination of the Gower distance and setting conditions on minimal number of
donors. In practice, imputation is often performed via a fall-though scenario,
where one first tries to estimate a model within a group, but if the group is
too small for a reliable estimate of model parameters, the group is enlarged by
combining similar groups, similar to the small-area estimation scenario.

SAE as a special case is well supported by \proglang{R} and other free
software. Methodology has for example been implemented in the \pkg{sae} package
of \cite{molina2015sae} and the \pkg{hbsae} package of
\cite{boonstra2022hbsae}.  Regarding imputation methodology, the CRAN task view
on missing data\footnote{https://cran.r-project.org/web/views/MissingData.html}
currently lists 203 \proglang{R} packages that support some form of estimating
missing data.  The \pkg{simputation} package \citep{loo2022simputation} seems
to be the only one that allows for some kind of fall-through scenario for
selecting methods, but it does not allow for dynamic grouping.

Summarizing, we see that on one hand there are many implementations available
for generic aggregation based on fixed groups. On the other hand there are
domain-specific implementation for methods where dynamic grouping of a set of
records plays a role. This paper presents a generic solution to
split-apply-combine aggregation where groups can be collapsed dynamically in
the form of \proglang{R} package \pkg{accumulate} \citep{loo2022accumulate}.

The \code{accumulate} package serves the use case where a user wishes to
compute aggregates for a certain grouping of records. However, if a certain
instance of a group does not meet user-defined quality requirements the set of
records is expanded by (recursively) collapsing the grouping according to a
user-defined scheme. For example, given some financial data on companies, one
wishes to compute the average profit to turnover ratio for each combination of
economic activity and size class. If for a certain combination of economic
activity and size class there are too few records for a reliable estimate, one
could drop size class and compute the average ratio over all records within a
certain economic activity. Or, one could choose to coarse-grain economic
activities by collapsing groups with activities that are deemed similar enough.

The package has been developed with the following design choices in mind.
First, the interface should be easy to learn for \proglang{R} users, and thus
should resemble existing popular interfaces where possible. Second, users
should be free to define any, possibly multi-step, collapsing scheme. Here, we
keep in mind that collapsing schemes may be constructed manually based on
domain knowledge and that users may want to experiment with several schemes
before deciding on a final solution. This calls for a certain separation of
concerns between defining collapsing schemes and applying them to data.  The
package should also support collapsing schemes that follow naturally from
hierarchical classification systems. Third, users should have the flexibility
to define any quality requirement on the grouping while common quality
requirements are supported out of the box. Common quality requirements include
a minimum number of records, or a minimum fraction of records without missing
values, or a minimum number of records with non-zero values. Finally, the
package should support simple outputs such as counts or averages, but also
compound objects such as the output of model estimates.

The rest of this paper is organized as follows.  In the next section we
visually explain dynamic grouping via a collapsing scheme and introduce the
running example that will be used in Section~\ref{sect:accumulate} which
introduces the \pkg{accumulate} package, the main interfaces and helper
functions. In Section~\ref{sect:accumulate} we also discuss the common case of
collapsing via a predifined hiearchical classification scheme.
Section~\ref{sect:example} demonstrates the package on a realistic synthetic
dataset. The case of complex outputs is also demonstrated in this Section.  In
Section~\ref{sect:algorithm} we derive the pseudocode that solves the problem
of aggregating with dynamically collapsing groups. We show that it is a precise
and straightforward generalization of the standard split-apply-combine problem
and analyze its time complexity. A summary and conlcusion follows in
Section~\ref{sect:conclusion}.

\section{Dynamic Grouping} \label{sect:dynamicgrouping}
In this Section the illustrate the concept of dynamic grouping with a minimal
worked example. This example is not very realistic, but it is constructed
to be simple enough so the whole procedure can be followed in detail. 

We consider a data set with three categorical variables $A$, $B$ and $B_1$, and
one numerical variable $Y$.  Variable $A$ has levels $\{1, 2,3\}$ and variable
$B$ is a hierarchical classification with levels $\{11,12,13,21,22\}$. Variable
$B_1$ is a coarse-graining of $B$: for each record the value for $B_1$ is the
first digit of $B$. Hence, $B_1$ has levels $\{1,2\}$. 

Our goal is to compute the mean $\mu_Y$ over $Y$, grouped by $A\times B$.  We
impose the condition that there must be at least three records in each group.
If a certain group $(a\in A,b\in B)$ has less than three records, we attempt
to compute the value for that group over records in $(a,b_1)$ where $b_1$ is
obtained by taking the first digit of $b$. If we then still have less than
three records, we take records of group $a$ to determine the value for $(a,b)$.

The tables in Figure~\ref{fig:example} illustrate the idea. The left table
represents the data set to be aggregated by $A$ and $B$.  The table on the
right represents the output. Colors indicate which data was used. 
\begin{figure}[H]
\centering
\begin{tabular}{cccl}
\multicolumn{4}{l}{Input Data}\\
\hline
$A$ & $B$ & $B_1$ & $Y$\\
\hline
\cellcolor{red!25}   1 &\cellcolor{red!25}   11 &                1       & \cellcolor{red!25}   1   \\
\cellcolor{red!25}   1 &\cellcolor{red!25}   11 &                1       & \cellcolor{red!25}   2   \\
\cellcolor{red!25}   1 &\cellcolor{red!25}   11 &                1       & \cellcolor{red!25}   3   \\
\cellcolor{green!25} 2 &                     12 & \cellcolor{green!25} 1 & \cellcolor{green!25} 4   \\
\cellcolor{green!25} 2 &                     12 & \cellcolor{green!25} 1 & \cellcolor{green!25} 5   \\
\cellcolor{green!25} 2 &                     13 & \cellcolor{green!25} 1 & \cellcolor{green!25} 6   \\
\cellcolor{blue!25}  3 &                     21 &                2       & \cellcolor{blue!25}  7   \\
\cellcolor{blue!25}  3 &                     22 &                2       & \cellcolor{blue!25}  8   \\
\cellcolor{blue!25}  3 &                     12 &                1       & \cellcolor{blue!25}  9   \\
\hline
\end{tabular}\hspace{1cm}\begin{tabular}{ccl}
\multicolumn{3}{l}{Output Aggregates}\\
\hline
$A\times B$ & \code{Level} & $\mu_Y$\\
\hline
\rowcolor{red!25}   1 11        & 0             & 2  \\
\rowcolor{green!25} 2 12        & 1             & 5  \\
\rowcolor{green!25} 2 13        & 1             & 5  \\
\rowcolor{blue!25}  3 21        & 2             & 8  \\
\rowcolor{blue!25}  3 22        & 2             & 8  \\
\rowcolor{blue!25}  3 12        & 2             & 8  \\
\hline
\end{tabular}
\label{fig:example}
\caption{Input data (left) with grouping variables $A$, $B$ and $B_1$, and
means of $Y$ per $A\times B$ after dynamic grouping (right).}
\end{figure}
The First row in the output represents group $(A=1,B=11)$. The collapsing level
is zero, which means that no collapsing was necessary. Indeed, in the data
table we see that there are three rows with $A=1$ and $B=11$ with $Y$ values
$1, 2$, and $3$, resulting in $\mu_Y=(1+2+3)/3=2$ for this group.

Next, we try to compute the total for group $(A=2,B=12)$ (in green) but find
that there are only two such rows. We now define a new group $(A=2,B_1=1)$ and
find that there are three records in that group so we get 
$\mu_Y=(4+5+6)/3=5$.  Similarly, there is only one record with $(A=2,B=13)$. 
Collapsing groups to $(A=2,B_1=1)$, yields again $\mu_y=5$.

Finally, for $(A=2,B=21)$ there is only a single record. Collapsing to
$(A=2,B_1=2)$ yields only two records, so we need to collapse further to $(A=2)$
and finally obtain three records.  This yields $\mu_Y=(7+8+9)/3=8$. Similarly
the groups $(A=3, B=22)$ and $(A=3,B=12)$ are collapsed to $(A=2)$.

\section[R Package accumulate]{\proglang{R Package} \pkg{accumulate} }\label{sect:accumulate}
Grouped aggregation with a fall-though scenario based on a collapsing scheme
requires a fair amount of specification by the user. Besides the data to be
aggregated, one needs to specify the method(s) of aggregation, the collapsing
scheme, and the condition to decide whether a subset is fit for aggregation or
a next collapse is necessary. There are two main functions in \pkg{accumulate}
that offer slightly different interfaces.
\begin{Code}
  accumulate(data, collapse, test, fun, ...)
  cumulate(data, collapse, test, ...)
\end{Code}
Here \texttt{data} is a data frame holding data to be aggregated;
\texttt{collapse} represents the collapse sequence (as a \class{formula} or a
\class{data frame}), and \texttt{test} is a function that accepts a subset of
\texttt{data} and returns a boolean that indicates whether a subset is suited
for aggregation or not. In \code{accumulate()}, the parameter \code{fun}
represents an aggregation function that is applied to every column of
\code{data}, and the ellipsis (\code{...}) is for arguments that are passed as
extra argument to \code{fun}. The interface of \code{accumulate()} is somewhat
similar to that of the \code{aggregate()} function in \proglang{R}.  In
\code{cumulate()}, the ellipsis is a sequence of comma-separated
\code{name = expression} pairs in the style of \code{summarise()} from the
\code{dplyr} package.

The output of both functions are of the same form. The columns of the output
data frame and can schematically be represented as follows. 
\begin{Code}
  [Grouping Variables, Collapse level, Output aggregates]
\end{Code}
The first columns represent the variables that define the output grouping, the
next column is an integer that indicates the level of collapsing used to
compute the aggregate (0 indicating no collapse), and the last set of columns
store the aggregates. Output aggregates may be of a simple data type
(\code{numeric}, \code{character}, \code{logical},$\ldots$) or of a composed
type such as the output of a linear model. The latter case is demonstrated in
Section~\ref{sect:complex}.

Both functions support two different interfaces for specifying collapsing
schemes through the \code{collapse} parameter. The first and most general is
the \class{formula} interface, which requires that the collapsing sequence is
represented as variables in the data set to be aggregated. The second is a
tabular interface where each row starts with the one of the lowest-level group
labels, and subsequent columns contain labels of courser groups.

\subsection{The Formula Interface}\label{sect:formula}
We will use the example of Section~\ref{sect:dynamicgrouping} to illustrate
the \class{formula} interface.
\begin{Schunk}
\begin{Sinput}
R> library("accumulate")
R> input <- data.frame(
+    A  = c( 1,  1,  1,  2,  2,  2,  3,  3,  3),
+    B  = c(11, 11, 11, 12, 12, 13, 21, 22, 12),
+    B1 = c( 1,  1,  1,  1,  1,  1,  2,  2,  1),
+    Y  = 1:9
+  )
R> cumulate(input, collapse = A * B ~ A * B1 + A
+          , test = function(d) nrow(d) >= 3, muY = mean(Y) )
\end{Sinput}
\begin{Soutput}
  A  B level muY
1 1 11     0   2
4 2 12     1   5
6 2 13     1   5
7 3 21     2   8
8 3 22     2   8
9 3 12     2   8
\end{Soutput}
\end{Schunk}
Consider the formula \code{A * B ~ A * B1 + A} in the call to \code{cumulate()}.
The left-hand-side \code{A * B} is the target output grouping. The
right-hand-side is to be interpreted as the collapsing sequence: if an instance
of \code{A * B} does not pass the test, then collapse to \code{A * B1}, and if that
does not pass the test collapse to \code{A}. If this final grouping also does
not pass the test, the result is \code{NA}.

Summarizing, the \class{formula} interface is always of the following form.
\begin{Code}
  Target grouping ~ Alternative1 + Alternative2 + ... + AlternativeN
\end{Code}

It is possible to get the same result with \code{accumulate()}. This will cause
summation over all variables that are not used in the formula object. In the
below example we also introduce the helper function \code{min_records()}.
\begin{Schunk}
\begin{Sinput}
R> input$Y2 <- 11:19
R> accumulate(input, collapse = A * B ~ A * B1 + A, 
+    test = min_records(3), fun = mean)
\end{Sinput}
\begin{Soutput}
  A  B level Y Y2
1 1 11     0 2 12
4 2 12     1 5 15
6 2 13     1 5 15
7 3 21     2 8 18
8 3 22     2 8 18
9 3 12     2 8 18
\end{Soutput}
\end{Schunk}
This means that for experimentation, users must be careful to exclude categorical
variables that are not used from the input data set. For example, if \code{B1}
is not used, we get the following.
\begin{Schunk}
\begin{Sinput}
R> accumulate(input[-3], collapse = A * B ~ A,
+    test = min_records(3), fun = mean)
\end{Sinput}
\begin{Soutput}
  A  B level Y Y2
1 1 11     0 2 12
4 2 12     1 5 15
6 2 13     1 5 15
7 3 21     1 8 18
8 3 22     1 8 18
9 3 12     1 8 18
\end{Soutput}
\end{Schunk}

The \class{formula} interface allows users to quickly experiment with different
collapsing schemes. It does require that all categorical variables have been
added to the data. As an alternative, one can use the data frame specification,
which allows for a further separation between defining the collapsing scheme
and the actual data processing.

\subsection{The Data Frame Interface} \label{sect:dfint}
The data frame interface is somewhat limited because it only allows for a
single grouping variable. The advantage however setting up a collapsing scheme
in the form of a table closely connects to domain knowledge and allows
fine-grained control on how groups are collapsed.

In order to use the data frame interface, the input dataset must include the
most fine-grained grouping variable. In the running example this is $A\times
B$, so we need to combine that into a single variable, and remove the other
ones.
\begin{Schunk}
\begin{Sinput}
R> input1 <- input
R> input1$AB <- paste(input$A, input$B, sep = "-")
R> input1 <- input1[-(1:3)]
R> input1
\end{Sinput}
\begin{Soutput}
  Y Y2   AB
1 1 11 1-11
2 2 12 1-11
3 3 13 1-11
4 4 14 2-12
5 5 15 2-12
6 6 16 2-13
7 7 17 3-21
8 8 18 3-22
9 9 19 3-12
\end{Soutput}
\end{Schunk}
We now define the collapsing scheme as follows.
\begin{Schunk}
\begin{Sinput}
R> csh <- data.frame(
+    AB  = c("1-11", "2-12", "2-13", "3-21", "3-22", "3-12"),
+    AB1 = c("1-1" , "2-1" , "2-1" , "3-2" , "3-2" , "3-1" ),
+    A   = c("1"   , "2"   , "2"   , "3"   , "3"   , "3"   ))
R> csh
\end{Sinput}
\begin{Soutput}
    AB AB1 A
1 1-11 1-1 1
2 2-12 2-1 2
3 2-13 2-1 2
4 3-21 3-2 3
5 3-22 3-2 3
6 3-12 3-1 3
\end{Soutput}
\end{Schunk}
In this data frame, two consecutive columns should be read as a child-parent
relation. For example, in the first collapsing step the groups defined by
\code{AB == "2-12"} and \code{AB == "2-13"} both collapse to \code{AB1 == "2-1"}.
In this artificial example the codes do not mean anything, but in realistic
cases where codes represent a (hierarchical) classification, domain experts
usually have a good grasp of which codes can be combined. 

The calls to \code{cumulate()} and \code{accumulate()} now look as follows.
\begin{Schunk}
\begin{Sinput}
R> accumulate(input1, collapse = csh, test = min_records(3), mean)
\end{Sinput}
\begin{Soutput}
    AB level Y Y2
1 1-11     0 2 12
4 2-12     1 5 15
6 2-13     1 5 15
7 3-21     2 8 18
8 3-22     2 8 18
9 3-12     2 8 18
\end{Soutput}
\begin{Sinput}
R> cumulate(input1, collapse = csh, test = min_records(3),
+    muY = mean(Y), muY2 = mean(Y2))
\end{Sinput}
\begin{Soutput}
    AB level muY muY2
1 1-11     0   2   12
4 2-12     1   5   15
6 2-13     1   5   15
7 3-21     2   8   18
8 3-22     2   8   18
9 3-12     2   8   18
\end{Soutput}
\end{Schunk}

\subsection{Specifying Tests}
The \code{test} parameter of \code{cumulate()} and \code{accumulate()} accepts a
function that takes a subset of the data and returns a boolean. For common
test conditions, including requiring a minimal number of records, or a minimal
number or fraction of complete records there are helper functions available.
\begin{center}
\begin{tabular}{lp{9cm}}
\code{min_records(n)}        & At least $n$ records.\\ 
\code{min_complete(n, vars)}  & At least $n$ records complete for variables \code{vars}.\\
\code{frac_complete(r, vars)} & At least $100r$\% complete records for variables \code{vars}.\\
\code{from_validator(v, ...)} & Construct a testing function from a \code{validator} object
                               of \proglang{R} package \pkg{validate}.\\
\end{tabular}
\end{center}
Second, for multiple, possibly complex requirements on variables users can
express conditions with the \pkg{validate} package \cite{loo2021data}.  The
\pkg{validate} packages offers a domain-specific language for expressing,
manipulating, and investigating conditions on datasets. It's core concept is a
list of `data validation rules' stored as a \class{validator} object. A
\class{validator} object is constructed with the eponymous function
\code{validator()}.  For example, to demand that there are at least 3 rows in a
group, and that there are at least three records where $Y\geq 2$ we create the
following ruleset.
\begin{Schunk}
\begin{Sinput}
R> library("validate")
R> rules <- validator(nrow(.) >= 3, sum(Y >= 2) >= 3)
R> rules
\end{Sinput}
\begin{Soutput}
Object of class 'validator' with 2 elements:
 V1: nrow(.) >= 3
 V2: sum(Y >= 2) >= 3
\end{Soutput}
\end{Schunk}
Here the \code{.} refers to the dataset as a whole, while rules that can be
evaluated within the dataset can be written as boolean \proglang{R}
expressions.

We will apply these conditions to the \code{input} dataset that was 
constructed in Section~\ref{sect:formula}. As a reminder we print
the first few records.
\begin{Schunk}
\begin{Sinput}
R> head(input, 4)
\end{Sinput}
\begin{Soutput}
  A  B B1 Y Y2
1 1 11  1 1 11
2 1 11  1 2 12
3 1 11  1 3 13
4 2 12  1 4 14
\end{Soutput}
\end{Schunk}
We use \code{A * B ~ A * B1 + B1} as collapsing scheme. The function
\code{from_validator} passes the requirements as a test function to
\code{accumulate()} (or \code{cumulate()}).
\begin{Schunk}
\begin{Sinput}
R> accumulate(input, collapse = A * B ~ A * B1 + B1,
+      test = from_validator(rules), fun = mean)
\end{Sinput}
\begin{Soutput}
  A  B level        Y       Y2
1 1 11     2 4.285714 14.28571
4 2 12     1 5.000000 15.00000
6 2 13     1 5.000000 15.00000
7 3 21    NA       NA       NA
8 3 22    NA       NA       NA
9 3 12     2 4.285714 14.28571
\end{Soutput}
\end{Schunk}
Note that for target groups $(A=3,B=21)$ and $(A=3,B=22)$ none of the available
collapsing levels lead to a group that satisfied all conditions. Therefore the
collapsing level and output variables are all missing (\code{NA}).

The third and most flexible way for users to express tests is to write a custom
testing function. The requirements are that it must work on any subset of a
data frame, including a dataset with zero rows. The previous example can
thus also be expressed as follows.
\begin{Schunk}
\begin{Sinput}
R> my_test <- function(d) nrow(d) >= 3 && sum(d$Y >= 2) >= 3
R> accumulate(input, collapse = A * B ~ A * B1 + B1,
+    test = my_test, fun = mean)
\end{Sinput}
\begin{Soutput}
  A  B level        Y       Y2
1 1 11     2 4.285714 14.28571
4 2 12     1 5.000000 15.00000
6 2 13     1 5.000000 15.00000
7 3 21    NA       NA       NA
8 3 22    NA       NA       NA
9 3 12     2 4.285714 14.28571
\end{Soutput}
\end{Schunk}

It is easy to overlook some edge cases when specifying test functions.
Recall that a test function is required to return \code{TRUE} or \code{FALSE},
regardless of the data circumstances. The only thing that a test function
can assume is that the received data set is a subset of records from the
dataset to be aggregated. As a service to the user, \pkg{accumulate} exports
a function that checks a test function against common edge cases, including
the occurrence of missing values, a dataset with zero rows, and the full
dataset. The function \code{smoke_test()} checks whether the output is \code{TRUE} or \code{FALSE}
under all circumstances and also reports errors, warnings, and messages.
It accepts a (realistic) dataset and a testing function and prints
test results to the console.
\begin{Schunk}
\begin{Sinput}
R> smoke_test(input, my_test)
\end{Sinput}
\begin{Soutput}
Test with full dataset and Y is NA for all records raised issues.

   NA detected in output (must be TRUE or FALSE)
\end{Soutput}
\end{Schunk}
By default only failing tests are printed. In this case our test function is
not robust against missing values for $Y$. This can be remedied by passing
\code{na.rm = TRUE} as parameter to \code{sum()} in the test function.
\begin{Schunk}
\begin{Sinput}
R> my_test1 <- function(d) nrow(d) >= 3 && sum(d$Y >= 2, na.rm = TRUE) >= 3 
R> smoke_test(input, my_test1)
\end{Sinput}
\begin{Soutput}

\end{Soutput}
\end{Schunk}
The smoke test is aimed at preventing complicated stack traces when errors occur
in a call to \code{accumulate()} or \code{cumulate()}.  Users should be aware
that it does not guarantee correctness of the results, only robustness against
certain edge cases.

\subsection{Balanced and Unbalanced Hierarchical Classifications}
Hierarchical classifications are abundant in (official) statistics.
They represent a classification of entities into non-overlapping, nested
groupings. Examples include the international standard industrial classification
of economic activities (ISIC, \citet{un2022isic}), the related 
statistical classification of economic activities in europe (NACE, \citet{eu2006nace})
and the Euroean skills, competences, qualifications and occupations classification
(ESCO, \citet{eu2022esco}).

Hierarchical classifications offer a natural mechanism for collapsing
fine-grained groupings into larger groups because of the parent-child
relationships. As an example consider a small piece of the NACE classification.

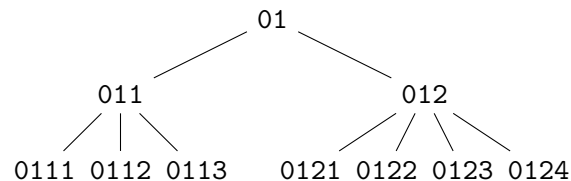
\begin{center}
\begin{tikzpicture}
\tikzstyle{level 1}=[level distance=10mm,sibling distance=40mm]
\tikzstyle{level 2}=[level distance=10mm,sibling distance=10mm]
  \node {\code{01}}
    child { node {\code{011}}
      child {node {\code{0111}}}
      child {node {\code{0112}}} 
      child {node {\code{0113}}} }
    child { node {\code{012}}
      child {node {\code{0121}}}
      child {node {\code{0122}}} 
      child {node {\code{0123}}} 
      child {node {\code{0124}}} };
\end{tikzpicture}
\end{center}
Here, the hierarchy suggests to collapse $\{0111,0112,0113\}$ into $\{011\}$
when needed, and similarly for the right branch. The second level of collapsing
would combine $\{012\}$ with $\{011\}$ into $\{01\}$. The \pkg{accumulate}
package comes with a helper function that creates the collapsing scheme from the
lowest-level digits.
\begin{Schunk}
\begin{Sinput}
R> nace <- c("0111", "0112", "0113", "0121", "0121", "0122", "0123", "0124")
R> csh_from_digits(nace, levels = 2)
\end{Sinput}
\begin{Soutput}
    A0  A1 A2
1 0111 011 01
2 0112 011 01
3 0113 011 01
4 0121 012 01
5 0121 012 01
6 0122 012 01
7 0123 012 01
8 0124 012 01
\end{Soutput}
\end{Schunk}
Here, the parameter \code{levels} determines how many collapsing steps will be
computed. Since all codes are prepended with zero, there is no need to collapse
$01$ any further. The output can be used as argument to the \code{collapse}
parameter of the \code{accumulate()} or \code{cumulate()} functions.

The situation becomes a little more involved when hierarchical classifications
form a tree such that the distance from leave to trunk is not the same for all
leaves (unbalanced tree). This occurs in practice, for example when local
organisations create an extra level of detail for some, but not all leaves.
Below is an example of such a situation.
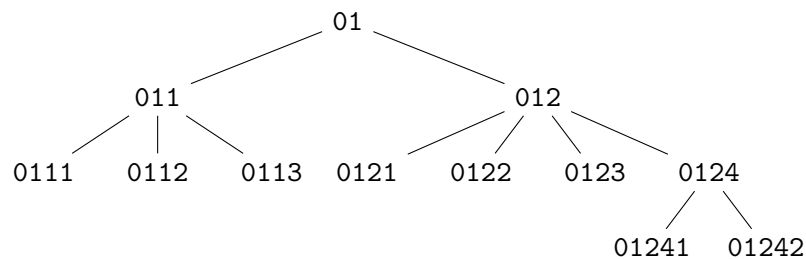
\begin{center}
\begin{tikzpicture}
\tikzstyle{level 1}=[level distance=10mm,sibling distance=50mm]
\tikzstyle{level 2}=[level distance=10mm,sibling distance=10mm]
\tikzstyle{level 2}=[level distance=10mm,sibling distance=15mm]
  \node {\code{01}}
    child { node {\code{011}}
      child {node {\code{0111}}}
      child {node {\code{0112}}} 
      child {node {\code{0113}}} }
    child { node {\code{012}}
      child {node {\code{0121}}}
      child {node {\code{0122}}} 
      child {node {\code{0123}}} 
      child {node {\code{0124}}
       child {node {\code{01241}}}
       child {node {\code{01242}}} } };
\end{tikzpicture}
\end{center}
In this case, not all leaves can be collapsed with the same number of
collapsing levels. This provides an issue for specifying the collapsing
sequence as it now depends on the leaf where you start how many collapsing
levels are possible. It also complicates interpretability of the result as the
collapsing level reported in the output, now means different things for
different target groups. The solution chosen in \pkg{accumulate} is to extend
the tree by making copies of the leaves that are not on the lowest level as
follows.
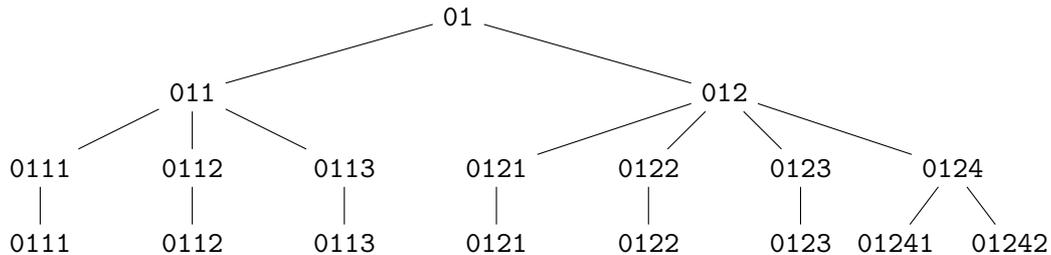
\begin{center}
\begin{tikzpicture}
\tikzstyle{level 1}=[level distance=10mm,sibling distance=70mm]
\tikzstyle{level 2}=[level distance=10mm,sibling distance=20mm]
\tikzstyle{level 3}=[level distance=10mm,sibling distance=15mm]
  \node {\code{01}}
    child { node {\code{011}}
      child {node {\code{0111}}
       child {node{\code{0111}}}} 
      child {node {\code{0112}} 
       child {node{\code{0112}}}} 
      child {node {\code{0113}}
       child {node{\code{0113}}}} } 
    child { node {\code{012}}
      child {node {\code{0121}}
       child {node {\code{0121}}}}
      child {node {\code{0122}}
       child {node{\code{0122}}}} 
      child {node {\code{0123}}
       child {node{\code{0123}}}} 
      child {node {\code{0124}}
       child {node {\code{01241}}}
       child {node {\code{01242}}} } };
\end{tikzpicture}
\end{center}
The tradeoff is that although there may be some extra calculations in the case
where a leaf is collapsed to itself. The gain is that the specification of the
calculation as well as the interpretation of the results are now uniform
across all hierarchical classifications. Again, using \code{csh_from_digits()}
deriving the collapsing scheme can be automated.
\begin{Schunk}
\begin{Sinput}
R> nace <- c("0111", "0112", "0113", "0121", "0122", "0123", "01241", "01242")
R> csh_from_digits(nace, levels = 3)
\end{Sinput}
\begin{Soutput}
     A0   A1  A2 A3
1  0111 0111 011 01
2  0112 0112 011 01
3  0113 0113 011 01
4  0121 0121 012 01
5  0122 0122 012 01
6  0123 0123 012 01
7 01241 0124 012 01
8 01242 0124 012 01
\end{Soutput}
\end{Schunk}

\section{Extensive Example: Economic Data} \label{sect:example}
In this Section we discuss three practical examples using a synthetic
dataset included with the package.
\begin{Schunk}
\begin{Sinput}
R> data("producers")
R> head(producers)
\end{Sinput}
\begin{Soutput}
   sbi size industrial trade other other_income  total
1 3410    8     151722  2135     0        -1775 152082
2 2840    7      50816    NA   158          949  59876
3 2752    5       4336    NA     0           36   4959
4 3120    6      18508    NA     0           80  20682
5 2524    7      21071     0     0          442  21513
6 3410    6      24220  1069     0          239  25528
\end{Soutput}
\end{Schunk}
This \code{producers} dataset contains synthetic data records of various income
sources for 1734 industrial produces. The records are
classified into a local version of the NACE classification called \code{sbi}
and into \code{size} classes with values in $\{5,6,7,8,9\}$. 

\subsection{Small Area Estimation} 
Small area estimation (SAE) is a collection of methods that are aimed at
estimating subpopulation parameters in cases where there the number of
observations in a subpopulation is so mall that direct estimation leads to
unacceptable estimation variance. Instead one may resort for example to
indirect estimation, meaning that one estimates parameters for a larger
subpopulation which are then used in the estimate for the target subpopulation.
Here, we shall be interested in estimating the average turnover from industrial
activities (\code{industrial}) by SBI and size class.

In the simplest case, where no auxiliary information is available or used, one
replaces the estimator of the mean over a subpopulation with the estimator of
the mean over a larger subpopulation that includes the target subpopulation
\citep[Section 3.2.1]{rao2015small}. If we assume that the dataset is obtained
by simple random sampling from the population, the mean can be estimated with
the sample mean. In this example we will demand that there are at least ten
records for which turnover has been measured. The collapsing scheme is given by
\code{sbi * size ~ sbi + sbi2 + sbi1} where \code{sbi2} and \code{sbi1} are
classification by respectively the first two SBI digits and the first SBI
digit. We first add those variables to the dataset.
\begin{Schunk}
\begin{Sinput}
R> producers$sbi2 <- substr(producers$sbi, 1, 2)
R> producers$sbi1 <- substr(producers$sbi, 1, 1)
R> head(producers, 3)
\end{Sinput}
\begin{Soutput}
   sbi size industrial trade other other_income  total sbi2 sbi1
1 3410    8     151722  2135     0        -1775 152082   34    3
2 2840    7      50816    NA   158          949  59876   28    2
3 2752    5       4336    NA     0           36   4959   27    2
\end{Soutput}
\end{Schunk}
Using \code{cumulate()} we obtain the means. 
\begin{Schunk}
\begin{Sinput}
R> a <- cumulate(producers,
+    collapse = sbi * size ~ sbi + sbi2 + sbi1,
+    test = min_complete(n = 10, vars = "industrial"),
+    mean_industrial = mean(industrial, na.rm = TRUE))
R> head(a,3)
\end{Sinput}
\begin{Soutput}
   sbi size level mean_industrial
1 3410    8     2        96157.67
2 2840    7     0        23160.25
3 2752    5     2        56966.33
\end{Soutput}
\end{Schunk}

In terms of SAE, the collapsing scheme expresses the assumption that estimates
on the level of respectively \code{sbi}, \code{sbi2} and \code{sbi1} introduce
an acceptable bias.

\subsection{Imputing Missing Values Using SAE and Ratio Imputation} \label{sect:ratio}
Our goal in this example is to impute missing values for the \code{industrial}
variable based on ratio imputation with \code{total} as predictor.  Ratio
imputation is a method where the imputed value $\hat{y}_i$ for variable $Y$ of
record $i$ is estimated as $\hat{y_i}=\hat{R}_d x_i$, where $\hat{R}_d$ is an
estimate of the ratio between $Y$ and an auxiliary variable $X$ in
subpopulation $d$. An unbiased estimate for $R_d$ is given by
$\hat{\bar{Y}}_d/\hat{\bar{X}}_d$ where $\hat{\bar{Y}}_d$ and $\hat{\bar{X}}_d$
are estimated subpopulation means. 

We use SAE to estimate the subpopulation ratios, and then use the \code{simputation}
package \citep{loo2022simputation} to impute the missing values. 
\begin{Schunk}
\begin{Sinput}
R> r <- cumulate(producers,
+    collapse = sbi * size ~ sbi + sbi2 + sbi1,
+    test = min_complete(n = 10, vars = "industrial"),
+    R = mean(industrial, na.rm = TRUE)/mean(total, na.rm = TRUE))
R> head(r,3)
\end{Sinput}
\begin{Soutput}
   sbi size level         R
1 3410    8     2 0.7450223
2 2840    7     0 0.8972639
3 2752    5     2 0.8725726
\end{Soutput}
\end{Schunk}
To impute the values, using \code{impute_proxy()} we need to merge the ratios
with the producers dataset (which automatically happens by SBI and size class).
\begin{Schunk}
\begin{Sinput}
R> library("simputation")
R> dat <- merge(producers, r)
R> dat <- impute_proxy(dat, industrial ~ R * total)
\end{Sinput}
\end{Schunk}
We can inspect the imputed values as follows.
\begin{Schunk}
\begin{Sinput}
R> iNA <- is.na(producers$industrial)
R> head(dat[iNA, c("sbi", "size", "level", "R", "industrial", "total")])
\end{Sinput}
\begin{Soutput}
       sbi size level         R industrial total
274   1581    5     0 0.9653030        830  1005
348   1581    6     0 0.9214068       5600  5600
392   1582    6     0 0.9521066       7969  8028
664  21122    7     2 0.9076063      45282 46149
1333  2840    6     0 0.9384657      17797 18186
1364  2851    6     0 0.9843794       7958  8315
\end{Soutput}
\end{Schunk}
From this, we get all the information to interpret the imputed values. For
example, we see that in record 664, the ratio was estimated after collapsing
the group $(\texttt{sbi},\texttt{size})=(21122,7)$ to $\texttt{sbi2}=21$ since
the collapse \code{level} equals 2.

\subsection{Random Nearest Neighbours Imputation with Collapsing Groups}
Nearest neighbor (NN) imputation is a donor imputation method where the
imputation value is copied from a record that is (randomly) chosen from a donor
pool \citep{andridge2010review}. In this example we use the grouping variables
in \code{producers} to define the donor pools. To prevent the same donor from
being used too often, it is not uncommon to demand a minimum number of records
in the donor pool. A collapsing scheme is one way of guaranteeing this, and
below we demonstrate how this problem can be expressed in \code{accumulate}.

We wish to impute the variable \code{trade} in the \code{producers} dataset
using donor imputation, where donors come from the same (\code{sbi},
\code{size}) combination. We wish to sample donor values from a group of at
least 5 donors. If this is not possible, we use the same fallback scenario as
in the previous Section.

We first define an `aggregation' function that takes a vector 
of donors and returns a non-empty sample.
\begin{Schunk}
\begin{Sinput}
R> random_element <- function(x) sample(x[ !is.na(x)], 1)
\end{Sinput}
\end{Schunk}
We will use \code{cumulate()} to ensure that there are at least five non-empty
values in \code{x} when \code{random_element()} is called. To make sure we
obtain a donor for each record, we add an identifying column \code{id} to use
as grouping variable.  
\begin{Schunk}
\begin{Sinput}
R> producers <- cbind(id = sprintf("ID%03d", seq_len(nrow(producers)))
+                    , producers)
R> set.seed(111)
R> imputations <- cumulate(producers
+                   , collapse = id ~ sbi * size + sbi + sbi2 + sbi1
+                   , test = min_complete(5,"trade")
+                   , donor_trade = random_element(trade))
R> head(imputations, 3)
\end{Sinput}
\begin{Soutput}
     id level donor_trade
1 ID001     3        2755
2 ID002     2        3293
3 ID003     3           0
\end{Soutput}
\end{Schunk}
To use the donor imputations, we merge the imputation candidates with the
original dataset and use \code{impute_proxy()} of the \pkg{simputation} package
for imputation.
\begin{Schunk}
\begin{Sinput}
R> imputed <- merge(producers, imputations) |>
+    impute_proxy(trade ~ donor_trade)
R> cols <- c(1:3, 9:10, 5)
R> head(producers[cols], 3)
\end{Sinput}
\begin{Soutput}
     id  sbi size sbi2 sbi1 trade
1 ID001 3410    8   34    3  2135
2 ID002 2840    7   28    2    NA
3 ID003 2752    5   27    2    NA
\end{Soutput}
\begin{Sinput}
R> head(imputed[c(cols, 11)], 3)
\end{Sinput}
\begin{Soutput}
     id  sbi size sbi2 sbi1 trade level
1 ID001 3410    8   34    3  2135     3
2 ID002 2840    7   28    2  3293     2
3 ID003 2752    5   27    2     0     3
\end{Soutput}
\end{Schunk}
The merge operation automatically merges on the \code{id} column, which also
adds the \code{level} column to the output.  The function \code{impute_proxy}
copies values from \code{donor_trade} into trade where \code{trade} is missing.
In the last expressions we only print the columns of interest.

\subsection{Computing Complex Aggregates} \label{sect:complex}
Until now, the aggregates have been simple (scalar) values.  With the
\code{cumulate()} function it is also possible to specify complex aggregates that
go beyond simple aggregates. Below, we estimate the following linear model
\begin{displaymath}
\texttt{total} = \beta_0 + \beta_i\texttt{industrial} + \varepsilon,
\end{displaymath}
demanding that there are at least 10 records where both predictor and predicted
variable are available.
\begin{Schunk}
\begin{Sinput}
R> r <- cumulate(producers,
+    collapse = sbi * size ~ sbi + sbi2 + sbi1,
+    test = min_complete(n = 10, vars = c("total","industrial")),
+    model = lm(total ~ industrial))
R> head(r,3)
\end{Sinput}
\begin{Soutput}
   sbi size level model
1 3410    8     2  <lm>
2 2840    7     0  <lm>
3 2752    5     2  <lm>
\end{Soutput}
\end{Schunk}
Here, the last column is a list of class \code{object_list}, where each
element is either an object of class \code{lm} or \code{NA}. Variables of class
\code{object_list} only differ from from a standard \proglang{R} list by their
print method.

\section{Formal Description and Algorithms} \label{sect:algorithm}
In this Section we give a formal description of the algorithm for aggregation
with dynamic grouping.  We start by giving an algorithm for ordinary
split-apply-combine to demonstrate how the algorithm must be generalized to
allow for a collapsing scheme.

\subsection{Split-Apply-Combine}
To analyse a data set group by group we need to specify a data set, a
way to split it into groups, and a function that takes a subset of data and
returns an aggregate. Let us introduce some notation for that.

Denote with $U$ a finite set, and let $\phi:2^U\to X$ be a function that
accepts a subset of $U$ and returns a value in some domain $X$.  Here, $U$
represents a data set, $2^U$ its power set, and $\phi$ an aggregating function.
We split $U$ into groups using the following notation. Let $A$ be finite set
that has no more elements than $U$, and let $f:U\onto A$ be a
surjective function that takes an element of $U$ and returns a value in $A$. We
can think of $A$ as a set of group labels, and $f$ as the function that assigns
a label to each element of $U$. This way, $f$ divides $U$ into non-overlapping
subsets. We say that $f:U\onto A$ is a \emph{partition} of $U$. We
also introduce the \emph{pullback along $f$}, $f^*:2^A\to 2^U$ defined as
\begin{equation*}
f^*(S) = \{u\in U|f(u)\in S\},
\end{equation*}
where $S$ is a subset of $A$ (See \emph{e.g.}, \citet[Section 1.4]{fong2019invitation}).

In this notation, any split-apply-combine operation can be computed with the
following algorithm.

\begin{algorithm}[H]
\caption{Split-Apply-Combine: $\textsc{SAC}(U,\phi,f)$}
\label{alg:sac}
\SetKwInOut{Input}{Input}\SetKwInOut{Output}{Output}
\Input{A finite set $U$, an aggregator $\phi: 2^U\to X$, and a partition
       $f:U\onto A$.}
\Output{$R$: the value of $\phi$ for every part of $U$ as a set of pairs
        $(a,x)\in A\times X$. }
$R = \{\}$\;
\For{$a\in A$}{
  $d = f^*(\{a\})$\tcp*{get subset of $U$}
  $R = R\cup \{(a,\phi(d))\}$\tcp*{aggregate and add to result}
}
\end{algorithm}
In this algorithm the output is collected in a set $R$ containing pairs from
$A\times X$: one pair for each element of $A$. (Incidently, the algorithm can be
summarized even shorter in this notation as $R=\cup_{a\in A}\{(a,(\phi\circ
f^*)(\{a\}))\}$, where $\circ$ denotes function composition).  

It is interesting to see how the elements $U$, $f$, and $\phi$ are implemented
in practice. Consider the signature of the \proglang{R}'s \code{aggregate()}
function (we skip arguments that are not important for the discussion).
\begin{Code}
  aggregate(x, by, FUN)
\end{Code}
Here, \code{x} is a data frame where each row represents an element of $U$. The
parameter \code{by} is a list of vectors of group labels, where each vector has
a length that equals the number of rows in \code{x}. So the function $f:U\onto
A$ is implemented by asking the user to make sure that the position of each
label in \code{by} corresponds to the correct row number in the data frame
\code{x}. The argument \code{FUN} (of class \class{function}) represents the
function $\phi$ that aggregates each subset of \code{x}. When the records in
\code{x} contain more than one variable, the aggregator is applied to each one
of them.  Here is an example of how a user might call this function from the
\proglang{R} prompt.
\begin{Schunk}
\begin{Sinput}
R> aggregate(iris[1:2], by = iris["Species"], FUN = mean)
\end{Sinput}
\begin{Soutput}
     Species Sepal.Length Sepal.Width
1     setosa        5.006       3.428
2 versicolor        5.936       2.770
3  virginica        6.588       2.974
\end{Soutput}
\end{Schunk}
Note that the correspondence in position of the \texttt{Species} label and the
record position is implemented by taking them from the same data frame. The
output also reveals in the first column the set $A$: each row corresponds to a
unique value in \texttt{Species} column.

\subsection{Split-Apply-Combine with Collapsing Groups}
\label{sect:saccg}
The goal of the algorithm is to compute a value for each part of a dataset,
possibly using values external to the part ---conditional to restrictions
placed on each part. The input of the algorithm consists again of a finite set
$U$ and an aggregation function $\phi$ that takes a subset of $U$ and returns a
value in some domain $X$.  Compared to Algorithm~\ref{alg:sac} two other inputs
are needed. First, a function must be defined that that checks whether a given
subset $d$ of $U$ is suitable for computing $\phi(d)$.  We will denote this
function $\beta: 2^U\to \mathbb{B}$, where
$\mathbb{B}=\{\texttt{True},\texttt{False}\}$.  Typical tests are checking
whether there are sufficient records available, or whether certain variables
have a low enough fraction of missing values. Second, we need a
\emph{collapsing scheme} $C$, defined as sequence of $n+1$ mappings
\begin{equation}
C\equiv U\xonto{f}A\xonto{f_1}A_1\xonto{f_2}\cdots\xonto{f_n}A_n.
\label{eq:collapsingsequence}
\end{equation}
A collapsing scheme is a sequence of partitions where each $f_i$ partitions its
domain in $|A_i|$ groups while $f$ partitions $U$ in $|A|$ groups.

Denote with $F_k:A\to A_k$, the function that  accepts a label in $A$ and
returns the corresponding label in $A_k$. In other words, $F_k$ is the
composition $f_k\circ f_{k-1}\circ\cdots \circ f_1$. Similarly we define the
pullback along $F_k$ as $F_k^* = f_1^*\circ f_2^*\circ\cdots\circ f_k^*$.  This
function accepts a set of labels in $A_k$ and returns all the labels in $A$
that are mapped to those labels via the collapsing sequence of
Equation~\ref{eq:collapsingsequence}.  With this notation we can define the
Algorithm for aggregation with dynamic grouping as follows. 

\begin{algorithm}[H]
\caption{Split-Apply-Combine with Collapsing Groups: $\textsc{SACCG}(U,\phi,\beta,C)$}
\label{alg:saccg}
\SetKwInOut{Input}{Input}\SetKwInOut{Output}{Output}
\Input{A finite set $U$, an aggregator $\phi: 2^U\to X$, a test function $\beta: 2^U\to \mathbb{B}$,
      and a collapsing sequence $C\equiv U\xonto{f}A\xonto{f_1}A_1\xonto{f_2}\cdots \xonto{f_n}  A_n$.}

\Output{$R$: the value of $\phi$ for every part of $U$, for which a suitable
collapsing group can be found, as a set of triples $(a,k,x)\in A\times
\underline{n}\times X $ where $\underline{n}=\{0,1,\ldots,n\}$.}

$R = \{\}$\;
\For{$a\in A$}{
  $i=0$ \tcp*{Initiate collapse level}
  $d = f^*(\{a\})$ \tcp*{Get subset of $U$}
  \While{$i<n \land \lnot\beta(d)$}{ \label{line:while}
    $i = i+1$ \tcp*{Increase collapse level} 
    $d = (f^*\circ F_i^*\circ F_i)(a)$ \tcp*{Collapse and get subset}
  }\label{line:endwhile}
  \If{$i<n \lor \beta(d)$}{ \label{line:cond}
    $R = R\cup \{(a, i,\phi(d))\}$\; \label{line:R}
  } 
}

\end{algorithm}

In this algorithm the collapsing level $i$ is increased until the test is
passed or the maximum collapsing level $n$ is reached
(Lines~\ref{line:while}-\ref{line:endwhile}) 
collapsing is determined dynamically by data 
algorithm also reports the collapsing level $i$ used.  The condition in
Line~\ref{line:cond} ensures that if no suitable dataset is found after the
whole collapsing sequence has been executed, then no answer is returned. This
means that in contrast with Algorithm~\ref{alg:sac} there is no guarantee that
a value for each member of $A$ will be found. For each member of $A$ where
an aggregate is computed, there is a triple $(a,i,\phi(d))$, where $a\in A$
is the label to which the value $\phi(d)$ pertains, and $i$ is the number of
collapses applied to reach a suitable dataset.

Comparing Algorithms~\ref{alg:sac} and~\ref{alg:saccg}, we see that the
standard split-apply-combine algorithm has worst-case runtime complexity
$O(|A|)$, as determined by counting applications of $f^*$. This is equal to the
algorithm's best case $\Omega(|A|)$. The split-apply-combine with collapsing
groups algorithm also has best case $\Omega(|A|)$ but has worst case $O(n|A|)$
(with $n$ the number of collapsing steps). In fact, the best case for
Algorithm~\ref{alg:saccg} is achieved by setting $\beta:d\mapsto\texttt{True}$.
In this case Algorithm~\ref{alg:saccg} reduces to Algorithm~\ref{alg:sac}. In
other words, we have
\begin{displaymath}
\textsc{SACCG}(U,\phi,\texttt{True},C) = \textsc{SAC}(U,\phi,f).
\end{displaymath}
To see this, observe that in Algorithm~\ref{alg:saccg} the condition in
line~\ref{line:while} is always \code{False} and the condition in
line~\ref{line:R} is always \code{True} in this case.

The worst case is achieved by setting $\beta: d\mapsto
\texttt{False}$. In that case the while loop in Line~\ref{line:while} is
iterated $n-1$ times (yielding total $n$ executions of $f^*$) while
Line~\ref{line:R} is never executed.

The analyses above leaves open the question of how the runtime depends on
application of the pullbacks in both algorithms. First note that there needs to
be no difference in applying $f^*$ or $f^*\circ F_i^*\circ F_i$: in practice a
collapsing scheme can be represented in tabular form just like $f$.  The
pullback then comes down to a lookup of records based on matching one or more
attributes with a set of attribute values. The time complexity of such
operations are typically reduced by proper preparation of the dataset. For
example, databases can be prepared to speed up certain often-used lookup
operations. In the implementation of the \code{accumulate} package the
pullback is implemented using standard join operations as
implemented by \proglang{R}'s standard \code{merge()} function.

\section{Summary and Conclusion}\label{sect:conclusion}
The \proglang{R} package \code{accumulate} introduced in this paper offers
convenient interfaces for computing grouped aggregates where the grouping is
dynamically determined based on user-defined conditions and a user-defined
group collapsing scheme. We demonstrated that these interfaces can be used in
several situations, including those where multiple grouping variables are used
, where complex collapsing schemes are applied, as well as situations where
multiple variables need to be aggregated over. Conditions on groups of records
are principally represented by a function, but the package includes several
convenience functions for defining conditions on record groups. It also
interfaces with the \pkg{validate} \proglang{R} package to facilitate cases
where mutliple test conditions must be met. Moreover, the package supports the
collection of complex aggregates, such as model outputs and offers several
service functions that aim to facilitate the definition of the collapsing
conditions.

The pseudocode underlying the package's main functions has been formally
analyzed and it is shown that aggregation with collapsing groups can be
interpreted as a precise generalization of standard grouped aggregation.  We
hope that this stimulates broader implementation of this algorithm in software
offering grouped aggregation.

\newpage
\bibliography{jss5097}

\begin{thebibliography}{21}
\newcommand{\enquote}[1]{``#1''}
\providecommand{\natexlab}[1]{#1}
\providecommand{\url}[1]{\texttt{#1}}
\providecommand{\urlprefix}{URL }
\expandafter\ifx\csname urlstyle\endcsname\relax
  \providecommand{\doi}[1]{doi:\discretionary{}{}{}#1}\else
  \providecommand{\doi}{doi:\discretionary{}{}{}\begingroup
  \urlstyle{rm}\Url}\fi
\providecommand{\eprint}[2][]{\url{#2}}

\bibitem[{Andridge and Little(2010)}]{andridge2010review}
Andridge RR, Little RJ (2010).
\newblock \enquote{A Review of Hot Deck Imputation for Survey Non-Response.}
\newblock \emph{International Statistical Review}, \textbf{78}(1), 40--64.

\bibitem[{Becker \emph{et~al.}(1988)Becker, Chambers, and
  Wilks}]{becker1988new}
Becker R, Chambers J, Wilks A (1988).
\newblock \enquote{The New \proglang{S} Language.}
\newblock \emph{Computer Science Series, Pacific Grove, CA}.

\bibitem[{Boonstra(2022)}]{boonstra2022hbsae}
Boonstra HJ (2022).
\newblock \emph{\pkg{hbsae}: Hierarchical Bayesian Small Area Estimation}.
\newblock \proglang{R} package version 1.2,
  \urlprefix\url{https://CRAN.R-project.org/package=hbsae}.

\bibitem[{Commision(2022)}]{eu2022esco}
Commision E (2022).
\newblock \enquote{European Skills, Competences, Qualifications, and
  Occupations.}
\newblock https://esco.ec.europa.eu.

\bibitem[{{Council of European Union}(2006)}]{eu2006nace}
{Council of European Union} (2006).
\newblock \enquote{Council Regulation ({EU}) No 1893/2006.}
\newblock
  \url{https://eur-lex.europa.eu/legal-content/EN/TXT/?uri=CELEX:32006R1893}.

\bibitem[{Dowle and Srinivasan(2024)}]{dowle2022datatable}
Dowle M, Srinivasan A (2024).
\newblock \emph{\pkg{data.table}: Extension of `\pkg{data.frame}'}.
\newblock \proglang{R} package version 1.15.4,
  \urlprefix\url{https://CRAN.R-project.org/package=data.table}.

\bibitem[{Eastwood(2023)}]{eastwood2022poorman}
Eastwood N (2023).
\newblock \emph{\pkg{poorman}: A Poor Man's Dependency Free Recreation of
  '\pkg{dplyr}'}.
\newblock \proglang{R} package version 0.2.7,
  \urlprefix\url{https://CRAN.R-project.org/package=poorman}.

\bibitem[{Fong and Spivak(2019)}]{fong2019invitation}
Fong B, Spivak DI (2019).
\newblock \emph{An Invitation to Applied Category Theory: Seven Sketches in
  Compositionality}.
\newblock Cambridge University Press.

\bibitem[{Kamiński \emph{et~al.}(2022)Kamiński, White, Bouchet-Valat,
  powerdistribution, Garborg, Quinn, Kornblith, cjprybol, Stukalov, Bates,
  Short, DuBois, Harris, Squire, Arslan, pdeffebach, Anthoff, Kleinschmidt,
  Noack, Shah, Mellnik, Arakaki, Mohapatra, Peter, Karpinski, Lin, Chagas,
  timema, ExpandingMan, and Oswald}]{kaminski2022dataframes}
Kamiński B, White JM, Bouchet-Valat M, powerdistribution, Garborg S, Quinn J,
  Kornblith S, cjprybol, Stukalov A, Bates D, Short T, DuBois C, Harris H,
  Squire K, Arslan A, pdeffebach, Anthoff D, Kleinschmidt D, Noack A, Shah VB,
  Mellnik A, Arakaki T, Mohapatra T, Peter, Karpinski S, Lin D, Chagas RAJ,
  timema, ExpandingMan, Oswald F (2022).
\newblock \emph{\pkg{JuliaData/DataFrames.jl}: v1.4.4}.

\bibitem[{Kowarik and Templ(2016)}]{kowarik2016imputation}
Kowarik A, Templ M (2016).
\newblock \enquote{Imputation with the \proglang{R} Package \pkg{VIM}.}
\newblock \emph{Journal of Statistical Software}, \textbf{74}(7), 1--16.
\newblock \doi{10.18637/jss.v074.i07}.

\bibitem[{Krantz(2024)}]{krantz2022collapse}
Krantz S (2024).
\newblock \emph{\pkg{collapse}: Advanced and Fast Data Transformation}.
\newblock \proglang{R} package version 2.0.13,
  \urlprefix\url{https://CRAN.R-project.org/package=collapse}.

\bibitem[{Molina and Marhuenda(2015)}]{molina2015sae}
Molina I, Marhuenda Y (2015).
\newblock \enquote{{sae}: An \proglang{R} Package for Small Area Estimation.}
\newblock \emph{The R Journal}, \textbf{7}(1), 81--98.
\newblock
  \urlprefix\url{https://journal.r-project.org/archive/2015/RJ-2015-007/RJ-2015-007.pdf}.

\bibitem[{Polars(2024)}]{polars2023}
Polars (2024).
\newblock \emph{Polars User Guide}.
\newblock \urlprefix\url{https://docs.pola.rs/}.

\bibitem[{\proglang{R} Core~Team(2024)}]{rcore}
\proglang{R} Core~Team (2024).
\newblock \emph{\proglang{R}: A Language and Environment for Statistical
  Computing}.
\newblock \proglang{R} Foundation for Statistical Computing, Vienna, Austria.
\newblock \urlprefix\url{https://www.R-project.org/}.

\bibitem[{Rao and Molina(2015)}]{rao2015small}
Rao JN, Molina I (2015).
\newblock \emph{Small Area Estimation}.
\newblock John Wiley \& Sons.

\bibitem[{{United Nations Statistical Division}(2008)}]{un2022isic}
{United Nations Statistical Division} (2008).
\newblock \enquote{International Standard Industrial Classification of Economic
  Activities (ISIC).}
\newblock \url{https://unstats.un.org/unsd/classifications/Econ/ISIC.cshtml}.

\bibitem[{{van der Loo}(2022)}]{loo2022simputation}
{van der Loo} M (2022).
\newblock \emph{\pkg{simputation}: Simple Imputation}.
\newblock \proglang{R} package version 0.2.8,
  \urlprefix\url{https://github.com/markvanderloo/simputation}.

\bibitem[{{van der Loo}(2023)}]{loo2022accumulate}
{van der Loo} M (2023).
\newblock \emph{\pkg{accumulate}: Split-Apply-Combine with Collapsing Groups}.
\newblock \proglang{R} package version 0.9.3.

\bibitem[{{van der Loo} and {de Jonge}(2021)}]{loo2021data}
{van der Loo} MPJ, {de Jonge} E (2021).
\newblock \enquote{Data Validation Infrastructure for \proglang{R}.}
\newblock \emph{Journal of Statistical Software}, \textbf{97}(10), 1--31.
\newblock \doi{10.18637/jss.v097.i10}.

\bibitem[{Wickham(2011)}]{wickham2011split}
Wickham H (2011).
\newblock \enquote{The Split-Apply-Combine Strategy for Data Analysis.}
\newblock \emph{Journal of Statistical Software}, \textbf{40}, 1--29.

\bibitem[{Wickham \emph{et~al.}(2024)Wickham, François, Henry, and
  Müller}]{wickham2022dplyr}
Wickham H, François R, Henry L, Müller K (2024).
\newblock \emph{\pkg{dplyr}: A Grammar of Data Manipulation}.
\newblock \proglang{R} package version 1.1.4,
  \urlprefix\url{https://CRAN.R-project.org/package=dplyr}.

\end{thebibliography}

\end{document}